\documentclass[12pt]{article}
\usepackage{amsfonts,amssymb,amsmath}
\usepackage[dvips]{epsfig}
\begin{document}
\title{\bf Robust construction of entangled coherent GHZ and W states in a cavity QED system}
\author{ N. Behzadi $^{a}$
\thanks{E-mail:n.behzadi@tabrizu.ac.ir}  ,
B. Ahansaz $^{b}$
\thanks{E-mail:bahramahansaz88@ms.tabrizu.ac.ir} ,
S. Kazemi $^{b}$
\thanks{E-mail:s.kazemirudsary89@ms.tabrizu.ac.ir}
\\ $^a${\small Research Institute for Fundamental Sciences,}
\\ $^b${\small Department of Theoretical Physics
and Astrophysics, Faculty of Physics,}
\\ {\small University of
Tabriz, Tabriz 51666-16471, Iran.}} \maketitle
\begin{abstract}
\noindent By exploiting a system of three distant cavities, we propose a scheme for constructing tripartite entangled coherent GHZ and W states which are robust due to the photon losses in the cavities. Each of cavities is doped with a semiconductor quantum dot. By the dynamics, the excitonic modes of quantum dots are enabled to exhibit entangled coherent GHZ and W states. Apart from the exciton losses, the master equation approach shows that when the populations of the field modes in the cavities are negligible the destruction of entanglement due to decoherence arises from photon losses, is effectively suppressed.
\\
\\
{\bf PACS Nos:} 03.65.Ud, 03.65.Fd
\\
{\bf Keywords:} GHZ-state and W-state, Coherent states, Tripartite entanglement, Decoherence, Master equation
\end{abstract}

\section{Introduction}
Cavity quantum electrodynamics (QED) provides a natural setting for distributed quantum information processing
(QIP) \cite{Cirac1}. One requirement of distributed QIP is the coupling of distant qubits in order to perform state transfer, entanglement generation, or quantum gate operations between separate nodes of the system. Coupled cavities not only can be considered as a tool for
observing of quantum cooperative phenomena in strongly
correlated many-body systems \cite{har} and also in observing strong coupling between photons and qubits inside the cavities \cite{dim, lam} but also has potential
applications in QIP \cite{ogd}. Recently, considerable theoretical efforts have been devoted to a class of coupled-cavity models that
promise to overcome the problem of individual addressability which is a difficult task in spin models \cite{li}. Furthermore, the interaction of a cavity and an atom can be engineered in such way that
the atom trapped in the cavity can have relatively long-lived energy levels which is suitable
for various QIP protocols such as entanglement generations \cite{uti}. Indeed, the system of high-Q cavities
and atoms, or specifically semiconductor quantum dots (QDs) as artificial atoms, in the strong interaction regime is one of experimentally realizable systems in
which the intrinsic quantum mechanical coupling dominates losses arisen due to dissipation
[6-8]. For example, in recent years, various
schemes based on cavity QED systems have
been proposed to generate entanglement between atoms or QDs
trapped in distant optical coupled cavities which would
be required for distributed quantum computing [9-13].

The discovery that tripartite entanglement can
provide a stronger violation of local realism \cite{gre, pan} than bipartite
entanglement has triggered a large research activity with the
aim to generate it and find its applications. For example, generating tripartite entanglement between three atoms trapped in three distant cavities connected by optical fibers via quantum Zeno
dynamics has been proposed in \cite{che, liw}. In \cite{lu}, deterministically
generating tripartite entangled states of distant
atoms based on the selective photon emission and absorption
processes, using three cavities linked by optical fibers,
has been discussed. It has
become clear that for the system shared by three parties there
are two inequivalent classes of entangled states, i.e., the
GHZ state and the W state \cite{vid} found applications in realizing quantum
information processing tasks \cite{kem}. Furthermore, the entangled coherent states (ECSs) \cite{san}
have emerged as a genuinely useful set of entangled states
having a prominent role, for instance, in quantum teleportation and quantum computing [22-24] so, in the same way, have did entangled coherent GHZ and W states \cite{an2, an3}.

In this work, we propose a protocol for generating tripartite entangled coherent GHZ and W states by exploiting three coupled-cavity QED which each of them is doped with a QD strongly coupled to it. We show that if excitonic mode of one of QDs is prepared in an even or odd coherent state and the other QDs and cavities are in their respective vacuum states, the dynamics of the system, at a characteristic time, layouts entanglement for the standard coherent states associated to excitonic modes of three distant noninteracting QDs. To evaluate the similarity of the generated entangled coherent states to coherent GHZ and W states quantitatively, we exploit GHZ state-based and W state-based entanglement witnesses (EWs) introduced in \cite{acin, bou}. Also, in this way, the process of constructing tripartite entanglement of coherent GHZ and W states is established in such a way that the effect of decoherence arisen almost from photon loss in cavities becomes negligible. In the resonance interaction between the field mode
of cavities and the respective excitonic mode of the QDs, the field mode extremely populates, which in turn, leads to decay of photons and therefore lose of coherency of the system \cite{dav, fon}. It is shown that, when the field mode is highly detuned with the excitonic mode, the establishment of entanglement can be satisfied without
populating the field mode in the cavities. Therefore, by the master equation approach, the proposed protocol for constructing entanglement, apart from the exciton losses, is robust to decoherence arisen from photon loss and thus, the efficient decoherence rate of the cavities is greatly prolonged.

This paper is laid out in the following way. In section 2, we describe the basic
properties of the scheme. In Section 3, we are going to describe the realization of the scheme and represent
a mechanism for dominating on the decoherence influencing the system to lose its coherency.
The paper is ended with a brief conclusion.

\section{ Hamiltonian and dynamics}
We consider three QDs trapped in the
three separated equidistance single-mode cavities placed at the
vertices of a equilateral triangle as depicted in Fig. 1. The size of each QD satisfies the condition $R\gg a_{B}$ (Bohr radius). It is assumed that in the quantum dots
there are a few electrons excited from
valanced-band to conduction-band and the excitation density
of the Coulomb correlated electron-hole pairs, excitons, in the
ground state for each quantum dot is low. This, in fact, indicates that the average number of excitons is no more than one for an effective area of the excitonic Bohr radius. Consequently, exciton
operators can be approximated with boson operators. Also, all
nonlinear terms including exciton-exciton interactions and the
phase-space filling effect can be neglected. It is also assumed that
the ground energy of the excitons in each quantum dot is the same.
The Hamiltonian Under the rotating wave approximation is given by
\begin{eqnarray}
&&\hspace{-7mm}\hat{H}=\hbar \omega_{c} \sum_{i=1}^{3}\hat{a}_{i}^{\dagger}\hat{a}_{i}+\hbar \omega_{e} \sum_{i=1}^{3}\hat{b}_{i}^{\dagger}\hat{b}_{i}+\hbar g\sum_{i=1}^{3}(\hat{b}_{i}^{\dagger}\hat{a}_{i}+\hat{b}_{i}\hat{a}_{i}^{\dagger})+\hbar c\sum_{i=1}^{3}(\hat{a}_{i}^{\dagger}\hat{a}_{i+1}+\hat{a}_{i+1}^{\dagger}\hat{a}_{i}),
\end{eqnarray}
where $a_{i}^{\dagger}$ ($a_{i}$) is the creation
(annihilation) operator for the ith ($4\equiv1\mathrm{mode}3$) cavity field mode with frequency $w_{c}$ and
$b_{i}^{\dagger}$ ($b_{i}$) is the creation(annihilation) operator for the ith excitonic mode with frequency $w_{e}$.
The coupling constants between quantum dot and cavity field are represented
by $g$ and the coupling strength between the
cavities is $c$, which in turn, depends strongly on both the
geometry of the cavities and the actual overlap between adjacent
cavities. By assuming $\hbar=1$ and $\omega_{e}=\omega_{c}-\Delta$, the Heisenberg equations of motion for the operators of cavities field and excitons lead to the first order differential matrix equation as follows
\begin{eqnarray}
\left(
  \begin{array}{c}
    \hat{\dot{a}}_{1} \\
     \hat{\dot{a}}_{2} \\
     \hat{\dot{a}}_{3} \\
     \hat{\dot{b}}_{1} \\
     \hat{\dot{b}}_{2} \\
     \hat{\dot{b}}_{3} \\
  \end{array}
\right)=-i\left(
          \begin{array}{cccccc}
            \omega_{c} & c & c & g & 0 & 0 \\
            c & \omega_{c} & c & 0 & g & 0 \\
            c & c & \omega_{c} & 0 & 0 & g \\
            g & 0 & 0 & \omega_{c}-\Delta & 0 & 0 \\
            0 & g & 0 & 0 & \omega_{c}-\Delta & 0 \\
            0 & 0 & g & 0 & 0 & \omega_{c}-\Delta \\
          \end{array}
        \right)\left(
                 \begin{array}{c}
                   \hat{a}_{1} \\
                   \hat{a}_{2} \\
                   \hat{a}_{3} \\
                   \hat{b}_{1} \\
                   \hat{b}_{2} \\
                   \hat{b}_{3} \\
                 \end{array}
               \right).
\end{eqnarray}
Solving this equation is very simple and so, in this way, $\hat{b}_{1}(t)$ is written as below and the other operators have been inserted in the appendix.
\begin{eqnarray}
\hat{b}_{1}(t)=u_{21}(t)\hat{a}_{1}(0)+u_{22}(t)\left(\hat{a}_{2}(0)+\hat{a}_{3}(0)\right)+v_{21}(t)\hat{b}_{1}(0)+v_{22}(t)(\hat{b}_{2}(0)+\hat{b}_{3}(0)),
\end{eqnarray}
where $u_{2j}(t)$ and $v_{2j}(t)$ for $j=1,2$ are denoted as
\begin{eqnarray}
\begin{array}{c}
  u_{21}(t)=\frac{ie^{-i\omega_{c} t}}{6}\left(e^{-i(c-\frac{\Delta}{2})t}(-A+\frac{(2c+\Delta)^2}{A})sin(\frac{At}{2})-2e^{i(\frac{c+\Delta}{2})t}(B-\frac{(c-\Delta)^2}{B})sin(\frac{Bt}{2})\right), \\\\
  u_{22}(t)=\frac{ie^{-i\omega_{c} t}}{6}\left(e^{-i(c-\frac{\Delta}{2})t}(-A+\frac{(2c+\Delta)^2}{A})sin(\frac{At}{2})+e^{i(\frac{c+\Delta}{2})t}(B-\frac{(c-\Delta)^2}{B})sin(\frac{Bt}{2})\right), \\\\
  v_{21}(t)=\frac{e^{-i\omega_{c} t}}{3}\left(e^{-i(c-\frac{\Delta}{2})t}(cos(\frac{At}{2})+\frac{i(2c+\Delta)}{A}sin(\frac{At}{2}))+2e^{i(\frac{c+\Delta}{2})t}(cos(\frac{Bt}{2})-\frac{i(c-\Delta)}{B}sin(\frac{Bt}{2}))\right), \\\\
  v_{22}(t)=\frac{e^{-i\omega_{c} t}}{3}\left(e^{-i(c-\frac{\Delta}{2})t}(cos(\frac{At}{2})+\frac{i(2c+\Delta)}{A}sin(\frac{At}{2}))-e^{i(\frac{c+\Delta}{2})t}(cos(\frac{Bt}{2})-\frac{i(c-\Delta)}{B}sin(\frac{Bt}{2}))\right), \end{array}
\end{eqnarray}
in which $A=\sqrt{4c^{2}+4c\Delta+\Delta^{2}+4g^{2}}$ and $B=\sqrt{c^{2}-2c\Delta+\Delta^{2}+4g^{2}}$.
We assume that the excitonic mode of the first QD  is prepared initially in a
superposition of two distinct coherent states $\left|\alpha_{1}\right\rangle$ and $\left|\alpha_{2}\right\rangle$ and the other excitonic and field modes are at their respective vacuum states so the initial state of the whole system is written as
\begin{eqnarray}
\left|\psi(0)\right\rangle=\frac{1}{\sqrt{N}}\left|0\right\rangle_{c_{1}}\left|0\right\rangle_{c_{2}}\left|0\right\rangle_{c_{3}}\left(\left|\alpha_{1}
\right\rangle+e^{i\theta}\left|\alpha_{2}\right\rangle\right)_{e_{1}}
\left|0\right\rangle_{e_{2}}\left|0\right\rangle_{e_{3}},
\end{eqnarray}
where
$N=\left(2+2cos(\theta+Im(\alpha^{\ast}_{1}\alpha_{2})\right)e^{-1/2(|\alpha_{1}|^{2}+|\alpha_{2}|^{2})+Re(\alpha^{\ast}_{1}\alpha_{2})}$
is the normalization coefficient and
$\left|\alpha\right\rangle:=e^{\alpha
\hat{a}^{\dagger}(0)-\alpha^{\ast}\hat{a}(0)}\left|0\right\rangle$
is defined as a standard coherent state.
The subscripts $c$ and $e$ stand for cavity and exciton modes respectively. Time evolution of the initial state is obtained as
\begin{eqnarray}
&&\hspace{-7.5mm}\left|\psi(t)\right\rangle=\frac{1}{\sqrt{N}}(\left|\alpha_{1}
u_{21}(t)\right\rangle_{c_{1}}\otimes\left|\alpha_{1}
u_{22}(t)\right\rangle_{c_{2}}\otimes\left|\alpha_{1}
u_{22}(t)\right\rangle_{c_{3}}\otimes\left|\alpha_{1}
v_{21}(t)\right\rangle_{e_{1}}\otimes\left|\alpha_{1}
v_{22}(t)\right\rangle_{e_{2}}\otimes\left|\alpha_{1}
v_{22}(t)\right\rangle_{e_{3}}\nonumber
\\
&&\hspace{-7.5mm}+e^{i\theta}\left|\alpha_{2} u_{21}(t)\right\rangle_{c_{1}}\otimes\left|\alpha_{2} u_{22}(t)\right\rangle_{c_{2}}\otimes\left|\alpha_{2} u_{22}(t)\right\rangle_{c_{3}}\otimes\left|\alpha_{2} v_{21}(t)\right\rangle_{e_{1}}\otimes\left|\alpha_{2} v_{22}(t)\right\rangle_{e_{2}}\otimes\left|\alpha_{2} v_{22}(t)\right\rangle_{e_{3}}).
\end{eqnarray}
The reduced density operator for the excitonic modes of three QDs is found as
\begin{eqnarray}
\begin{array}{c}
  \hat{\rho}_{e}(\theta, \alpha_{1}, \alpha_{2};t)=\frac{1}{N}(\left|\alpha_{1} v_{21}(t), \alpha_{1} v_{22}(t), \alpha_{1} v_{22}(t)\right\rangle\left\langle \alpha_{1} v_{21}(t), \alpha_{1} v_{22}(t), \alpha_{1} v_{22}(t)\right| \\\\
  +\left|\alpha_{2} v_{21}(t),\alpha_{2} v_{22}(t),\alpha_{2} v_{22}(t)\right\rangle\left\langle \alpha_{2} v_{21}(t),\alpha_{2} v_{22}(t),\alpha_{2} v_{22}(t)\right| \\\\
  +q_{1}(t)q_{2}(t)^{2}e^{-i\theta}\left|\alpha_{1} v_{21}(t), \alpha_{1} v_{22}(t), \alpha_{1} v_{22}(t)\right\rangle\left\langle \alpha_{2} v_{21}(t),\alpha_{2} v_{22}(t),\alpha_{2} v_{22}(t)\right| \\\\
  +q_{1}^{\ast}(t)q_{2}^{\ast}(t)^{2}e^{i\theta}\left|\alpha_{2} v_{21}(t),\alpha_{2} v_{22}(t),\alpha_{2} v_{22}(t)\right\rangle\left\langle \alpha_{1} v_{21}(t), \alpha_{1} v_{22}(t), \alpha_{1} v_{22}(t)\right|),
\end{array}
\end{eqnarray}
where $q_{1}(t)=e^{-1/2(|\alpha_{1}|^{2}+|\alpha_{2}|^{2}-2\alpha_{2}^{\ast}\alpha_{1})|u_{21}|^{2}}$ and $q_{2}(t)=e^{-1/2(|\alpha_{1}|^{2}+|\alpha_{2}|^{2}-2\alpha_{2}^{\ast}\alpha_{1})|u_{22}|^{2}}$. It is evident that, by the dynamics, the reduced density operator for the excitonic modes becomes, in general, mix.

\section{Entanglement}
Let us consider $\alpha_{1}=-\alpha_{2}=\alpha$, the density matrix $\hat{\rho}(\theta, \alpha; t)$ in (7) becomes as
\begin{eqnarray}
\begin{array}{c}
  \hat{\rho}(\theta, \alpha; t)=\frac{1}{N}(\left|\alpha v_{21}(t), \alpha v_{22}(t), \alpha v_{22}(t)\right\rangle\left\langle \alpha v_{21}(t), \alpha v_{22}(t), \alpha v_{22}(t)\right| \\\\
  +\left|-\alpha v_{21}(t), -\alpha v_{22}(t), -\alpha v_{22}(t)\right\rangle\left\langle -\alpha v_{21}(t), -\alpha v_{22}(t), -\alpha v_{22}(t)\right| \\\\
  +q_{1}(t)q_{2}(t)^{2}(e^{-i\theta}\left|\alpha v_{21}(t), \alpha v_{22}(t), \alpha v_{22}(t)\right\rangle\left\langle -\alpha v_{21}(t), -\alpha v_{22}(t), -\alpha v_{22}(t)\right| \\\\
  +e^{i\theta}\left|-\alpha v_{21}(t), -\alpha v_{22}(t), -\alpha v_{22}(t)\right\rangle\left\langle \alpha v_{21}(t), \alpha v_{22}(t), \alpha v_{22}(t)\right|)),
\end{array}
\end{eqnarray}
One can always rebuild two orthogonal and normalized states as basis of the two-dimensional Hilbert space using original two nonorthogonal coherent states, i.e.
\begin{eqnarray}
\begin{array}{c}
  \left|\alpha v_{21}(t)\right\rangle:=\left|0\right\rangle,\quad \left|-\alpha v_{21}(t)\right\rangle:=p_{1}(t)\left|0\right\rangle+\sqrt{1-p_{1}^{2}(t)}\left|1\right\rangle,\\\\
  \left|\alpha v_{22}(t)\right\rangle:=\left|0\right\rangle,\quad \left|-\alpha v_{22}(t)\right\rangle:=p_{2}(t)\left|0\right\rangle+\sqrt{1-p_{2}^{2}(t)}\left|1\right\rangle,
\end{array}
\end{eqnarray}
where $\left|0\right\rangle$ and $\left|1\right\rangle$ are orthogonal basis and $p_{1}(t)=e^{-2|\alpha|^{2}|v_{21}(t)|^{2}}$ and
$p_{2}(t)=e^{-2|\alpha|^{2}|v_{22}(t)|^{2}}$. By rewritten the nonorthogonal basis of the density matrix $\hat{\rho}(\theta, \alpha; t)$ in terms
of the orthogonal ones $\left|0\right\rangle$ and
$\left|1\right\rangle$ the density matrix $\hat{\rho}(\theta, \alpha; t)$ is encoded
as a three qubit quantum state, so the detect of generated entanglement between excitonic modes in three QDs becomes as the detection of entanglement in three-qubit quantum states. To this aim, we exploit EWs introduced in \cite{acin, bou}. At first, we consider the GHZ state-based EW as follows
\begin{eqnarray}
\hat{W}=\frac{1}{2}1\otimes1\otimes1-\left|GHZ\right\rangle\left\langle
GHZ\right|.
\end{eqnarray}
where
$\left|GHZ\right\rangle=\frac{1}{\sqrt{2}}(\left|000\right\rangle+\left|111\right\rangle)$
is the famous GHZ-state and 1/2 is the maximal squared overlap
between $|GHZ\rangle$ and the convex set of biseparable states. The expectation values of $\hat{W}$ with respect to
$\hat{\rho}(\theta, \alpha; t)$, after encoding as a three qubit
density matrix, is as
\begin{eqnarray}
\begin{array}{c}
  E_{_{GHZ}}\left(\hat{\rho}(\theta, \alpha;
t)\right)\equiv Tr(\hat{W}\hat{\rho}(\theta, \alpha;
t))=\frac{1}{2}-F(|GHZ\rangle\langle GHZ|, \hat{\rho}(\theta, \alpha; t))^{2}\\\\
  \end{array}
\end{eqnarray}
where $F(|GHZ\rangle\langle GHZ|, \hat{\rho}(\theta, \alpha;
t))=\sqrt{\langle GHZ|\hat{\rho}(\theta, \alpha; t)|GHZ\rangle}$ is
the fidelity between $\hat{\rho}(\theta, \alpha; t)$ and the
GHZ-state. It appears that for density operator
$\hat{\rho}(\theta=0, 2; t)$ the expectation values of the witness operator
$\hat{W}$ for some times $t^{\ast}$ becomes very close to -1/2, that is, $F(|GHZ\rangle\langle
GHZ|, \hat{\rho}(0, 2; t^{\ast}))\simeq1$. This shows that when
the value of $\alpha$ is sufficiently large (for example is 2), the state
$\hat{\rho}(0, 2; t)$ of the three excitonic modes approaches to the GHZ-state as shown in Fig. 2 and Fig. 3, and so the entanglement of $\hat{\rho}(0, 2; t)$ can be considered as entanglement of coherent GHZ-state \cite{bar}.
Also, we can obtain the average number of photons in the cavity field modes as follows
\begin{eqnarray}
\langle n_{c}\rangle=\frac{(1-e^{-2|\alpha|^{2}})|\alpha|^{2}(|u_{21}(t)|^{2}+2|u_{22}(t)|^{2})}{1+e^{-2|\alpha|^{2}}}, \end{eqnarray}
where it has been depicted in Fig. 2 and Fig. 3, for resonance interaction between photon and exciton and nonresonance one respectively.

In the next step, we consider W state-based EW \cite{acin, bou} as
\begin{eqnarray}
\hat{W}'=\frac{2}{3}1\otimes1\otimes1-\left|W\right\rangle\left\langle
W\right|,
\end{eqnarray}
where
$|W\rangle=\frac{1}{\sqrt{3}}(|001\rangle+|010\rangle+|100\rangle)$
is the W-state and 2/3 corresponds to the maximal squared overlap
between $\left|W\right\rangle$ and the set of biseparable states B. Expectation values of the witness operator $\hat{W}'$
with respect to the $\hat{\rho}(\theta, \alpha; t)$ yields as
\begin{eqnarray}
E_{_{W}}\left(\hat{\rho}(\theta, \alpha;
t)\right)\equiv Tr(\hat{W}'\hat{\rho}(\theta, \alpha;
t))=\frac{2}{3}-F(|W\rangle\langle W|, \hat{\rho}(\theta, \alpha;
t))^{2},
\end{eqnarray}
where $F(|W\rangle\langle W|, \hat{\rho}(\theta, \alpha;
t))=\sqrt{\langle W|\hat{\rho}(\theta, \alpha; t)|W\rangle}$ is the
fidelity between $\hat{\rho}(\theta, \alpha; t)$ and the W-state. For arbitrarily small values of $\alpha$ such as 0.01, if we choose
$\theta=\pi$ the expectation values of $\hat{W}'$ with respect to
$\hat{\rho}(\pi, 0.01; t))$ for some times $t^{\ast}$ becomes $-1/3$, as shown in
Fig. 4 and Fig. 5, which means that $F(|W\rangle\langle W|,
\hat{\rho}(\pi, 0.01; t^{\ast}))\simeq1$. Consequently, by the dynamics
of the system, the excitonic modes of three QDs construct entangled coherent W-state, the other type
of tripartite entangled coherent state which are not equivalent to the
GHZ-state \cite{bar}.
Also, it can be obtained the average number of photons in the field modes of cavities as follows
\begin{eqnarray}
\langle n_{c}\rangle=\frac{(1+e^{-2|\alpha|^{2}})|\alpha|^{2}(|u_{21}(t)|^{2}+2|u_{22}(t)|^{2})}{1-e^{-2|\alpha|^{2}}},
\end{eqnarray}
where for resonance and nonresonance interaction regimes, has been sketched in the Fig. 4 and Fig. 5 respectively. On the other hand, in the presence of resonance interaction between photons and excitons, for both cases of construction of entangled coherent GHZ and W states, the field modes of cavities highly populated as shown in the Fig. 2 and Fig. 4. This means that in the presence of interaction between the cavity system and environment, it is evident from
\cite{dav, fon} that: the larger the average number of photons inside the cavities, the faster will the
coherence decay. Therefore, decoherency aspects of the system dominates to coherency one and therefore the suggested protocol for generation of entanglement is not utilizable. One of the essential improvement in robust construction of  entanglement using a system of coupled cavities is the prevention of populating
of the field mode in the cavities \cite{ogd}. To this end, we see that in the nonresonance interaction regime ($\Delta\neq0$), the average number of photons in the field mode of the cavities can become a vanishing amount as depicted in Fig. 3 and Fig. 5. Therefore, in the nonresonance regime, during the process of generating entanglement between excitons of QDs in the cavities which takes place slowly rather than to the resonance case, the field mode of each cavity is approximately at the respective vacuum state \cite{ogd, zhe, maj}. In another words, generating entanglement between the excitonic modes in distant QDs, can be done by negligible populations of the field modes in the coupled-cavity system, protecting against decoherence via related decays of cavities.

\section{Explicit evaluation of decoherence effect on the entanglement}
Decoherences and environmental losses are important effects
in quantum information processing [18]. In the presence of Decoherences, the system due to the coupling to its environment is open and its dynamics, consists of two coherent and incoherent parts, can be described within the framework of master equations of Lindblad form \cite{gar}. The system considered in this paper (Fig. 1.) can be  subjected to several dissipative processes as sources of decoherence, such as cavities
losses at decay rates $\gamma_{c_{i}}$s ($i=1, 2, 3$), which in turn, arise due to interaction of the field mode in each cavity with a common reservoir \cite{lou}. Also, another dissipations are related to the interaction of excitons in the QDs with the same reservoir at rates $\gamma_{e_{i}}$s ($i=1, 2, 3$). Since the three cavities are equivalent and so the QDs, we take $\gamma_{c}:=\gamma_{c_{1}}=\gamma_{c_{2}}=\gamma_{c_{3}}$ and $\gamma_{e}:=\gamma_{e_{1}}=\gamma_{e_{2}}=\gamma_{e_{3}}$. Hence, in the schrodinger picture, the time evolution of the whole system is described by the following master equation for the density operator $\hat{\rho}$ as
\begin{eqnarray}
\begin{array}{c}
  \dot{\hat{\rho}}=i[\hat{\rho}, \hat{H}]+\sum_{i=1}^{3}\left(2\hat{L}_{c_{i}}\rho\hat{L}^{\dagger}_{c_{i}}-
\hat{L}_{c_{i}}\hat{L}^{\dagger}_{c_{i}}\hat{\rho}-\hat{\rho}\hat{L}_{c_{i}}\hat{L}^{\dagger}_{c_{i}}\right)
 \\\\
  +\sum_{i=1}^{3}\left(2\hat{L}_{e_{i}}\rho\hat{L}^{\dagger}_{e_{i}}-
\hat{L}_{e_{i}}\hat{L}^{\dagger}_{e_{i}}\hat{\rho}-\hat{\rho}\hat{L}_{e_{i}}\hat{L}^{\dagger}_{e_{i}}\right), \end{array}
\end{eqnarray}
where $\hat{L}_{c_{i}}=\sqrt{\gamma_{c}}\ \hat{a}_{i}$ and $\hat{L}_{e_{i}}=\sqrt{\gamma_{e}}\ \hat{b}_{i}$ with $i=1, 2, 3$, are the Lindblad operators corresponding to the cavity field modes and the excitons of QDs and $\hat{H}$ is the Hamiltonian of the system introduced in Eq. 1. If we transfer the  description into the Heisenberg picture, the time evolution of a typical operator of the system such as $\hat{O}$, is obtained by the following equation \cite{gar, ple},
\begin{eqnarray}
\begin{array}{c}
  \dot{\hat{O}}=-i[\hat{O}, \hat{H}]+\sum_{i=1}^{3}\left(2\hat{L}^{\dagger}_{c_{i}}\hat{O}\hat{L}_{c_{i}}-
\hat{L}^{\dagger}_{c_{i}}\hat{L}_{c_{i}}\hat{O}-\hat{O}\hat{L}^{\dagger}_{c_{i}}\hat{L}_{c_{i}}\right)
 \\\\
  +\sum_{i=1}^{3}\left(2\hat{L}^{\dagger}_{e_{i}}\hat{O}\hat{L}_{e_{i}}-
\hat{L}^{\dagger}_{e_{i}}\hat{L}_{e_{i}}\hat{O}-\hat{O}\hat{L}^{\dagger}_{e_{i}}\hat{L}_{e_{i}}\right). \end{array}
\end{eqnarray}
In the right hand side of the Eq. 17, the first term represents the coherent or unitary evolution of the operator and the next terms show the incoherent or dissipative ones. Eq. 17, for the operators $\hat{a}_{i}$s and $\hat{b}_{i}$s ($i=1, 2, 3$), gives the set of first order differential equations similar to the Eq. 2, except that we should consider $\omega_{c}\rightarrow \omega_{c}-i\gamma_{c}$ and $\Delta\rightarrow \Delta+i(\gamma_{e}-\gamma_{c})$. By considering these replacements, the solutions of the set of differential equations for the operators $\hat{a}_{i}(t)$s and $\hat{b}_{i}(t)$s ($i=1, 2, 3$) of the open system,
are obtain analytically as obtained for its closed counterpart discussed in the section 2. Now by considering the initial state in Eq. 5, in the presence of related dissipations, the time evolution of the reduced density matrix for excitonic modes of three QDs namely $\hat{\rho}_{_{diss}}(\theta, \alpha; t)$, in similar way as for $\hat{\rho}(\theta, \alpha; t)$ in Eq. 8, is obtained. The expectation values of the witness operator $\hat{W}$ with respect to the density matrix $\hat{\rho}_{_{diss}}(0, 2; t)$, that is, $E_{_{GHZ}}\left(\hat{\rho}_{_{diss}}(0, 2;
t)\right)$ with $\gamma_{e}=0.001$ and $\gamma_{c}=0$, at the resonance and nonresonance cases have been shown in Fig. 6 and Fig. 7 respectively. As it is evident, the existence of losses of excitons in QDs destroys the entanglement of coherent GHZ-states generated in both resonance and nonresonance cases. However, when the photon losses in the cavities are included (for example with  $\gamma_{c}=0.05$), the destruction rates of the entanglement for the resonance and nonresonance cases are considerably different as shown in Fig. 8 and Fig. 9. It is obvious that the entanglement in the nonresonance case is more robust due to the photon losses in the cavities than the resonance case. In fact, as discussed in the previous section, the negligible populations of the field modes in the cavities at the nonresonance regime suppresses the destruction of the entanglement due to the photon loesses. On the other hand, similar observations are obtained in constructing entangled coherent W-state by considering the same dissipation processes as it can be seen in the Fig. 10, Fig. 11, Fig. 12, Fig. 13.

\section{Conclusions}
We have presented a protocol for robust construction of entangled coherent GHZ and W states between the excitonic modes of three QDs trapped in coupled-cavity system. This protocol
utilizes the cavity field, induced nonlocal interaction, to couple three QDs for establishing tripartite entanglement between created excitons. As illustrated by the master equation approach, the negligible populations of the field modes in the cavity system suppress efficiently the effect of photon losses on the generated entanglement.

\newpage
\vspace{1cm} \setcounter{section}{0}
 \setcounter{equation}{0}
 \renewcommand{\theequation}{\arabic{equation}}
{\Large{Appendix:}}\\
The solutions of (2) for $\hat{a}_{2}(t)$, $\hat{a}_{3}(t)$ and
$\hat{b}_{1}(t)$, $\hat{b}_{2}(t)$ and $\hat{b}_{3}(t)$ are given as
follow:
\begin{eqnarray}
\hat{a}_{1}(t)=u_{11}(t)\hat{a}_{1}(0)+u_{12}(t)(\hat{a}_{2}(0)+\hat{a}_{3}(0))+v_{11}(t)\hat{b}_{1}(0)+v_{12}(t)(\hat{b}_{2}(0)+\hat{b}_{3}(0)),
\end{eqnarray}
\begin{eqnarray}
\hat{a}_{2}(t)=u_{11}(t)\hat{a}_{2}(0)+u_{12}(t)(\hat{a}_{1}(0)+\hat{a}_{3}(0))+v_{11}(t)\hat{b}_{2}(0)+v_{12}(t)(\hat{b}_{1}(0)+\hat{b}_{3}(0)),
\end{eqnarray}
\begin{eqnarray}
\hat{a}_{3}(t)=u_{11}(t)\hat{a}_{3}(0)+u_{12}(t)(\hat{a}_{1}(0)+\hat{a}_{2}(0))+v_{11}(t)\hat{b}_{3}(0)+v_{12}(t)(\hat{b}_{1}(0)+\hat{b}_{2}(0)),
\end{eqnarray}
\begin{eqnarray}
\hat{b}_{2}(t)=u_{21}(t)\hat{a}_{2}(0)+u_{22}(t)(\hat{a}_{1}(0)+\hat{a}_{3}(0))+v_{21}(t)\hat{b}_{2}(0)+v_{22}(t)(\hat{b}_{1}(0)+\hat{b}_{3}(0)),
\end{eqnarray}
\begin{eqnarray}
\hat{b}_{3}(t)=u_{21}(t)\hat{a}_{3}(0)+u_{22}(t)(\hat{a}_{1}(0)+\hat{a}_{2}(0))+v_{21}(t)\hat{b}_{3}(0)+v_{22}(t)(\hat{b}_{1}(0)+\hat{b}_{2}(0)),
\end{eqnarray}
where $u_{1j}(t)$ and $v_{1j}(t)$ for $j=1,2$ are denoted as
\begin{eqnarray}
\begin{array}{c}
  u_{11}(t)=\frac{e^{-i\omega_{c} t}}{3}\left(e^{-i(c-\frac{\Delta}{2})t}(cos(\frac{At}{2})-\frac{i(2c+\Delta)}{A}sin(\frac{At}{2}))+2e^{i(\frac{c+\Delta}{2})t}(cos(\frac{Bt}{2})+\frac{i(c-\Delta)}{B}sin(\frac{Bt}{2}))\right) \\\\
  u_{12}(t)=\frac{e^{-i\omega_{c} t}}{3}\left(e^{-i(c-\frac{\Delta}{2})t}(cos(\frac{At}{2})-\frac{i(2c+\Delta)}{A}sin(\frac{At}{2}))-e^{i(\frac{c+\Delta}{2})t}(cos(\frac{Bt}{2})+\frac{i(c-\Delta)}{B}sin(\frac{Bt}{2}))\right) \\\\
  v_{11}(t)=\frac{ie^{-i\omega_{c} t}}{6 g}\left(e^{-i(c-\frac{\Delta}{2})t}(-A+\frac{(2c+\Delta)^2}{A})sin(\frac{At}{2})-2e^{i(\frac{c+\Delta}{2})t}(B-\frac{(c-\Delta)^2}{B})sin(\frac{Bt}{2})\right) \\\\
  v_{12}(t)=\frac{ie^{-i\omega_{c} t}}{6 g}\left(e^{-i(c-\frac{\Delta}{2})t}(-A+\frac{(2c+\Delta)^2}{A})sin(\frac{At}{2})+e^{i(\frac{c+\Delta}{2})t}(B-\frac{(c-\Delta)^2}{B})sin(\frac{Bt}{2})\right) \end{array}
\end{eqnarray}

\newpage
\textbf{Figure Captions}
\itemize{}
\item Fig. 1. Three optical cavities each of which is doped with a QD, coupled to each other through photon hopping.
\begin{figure}
\centering
\includegraphics[width=445 pt]{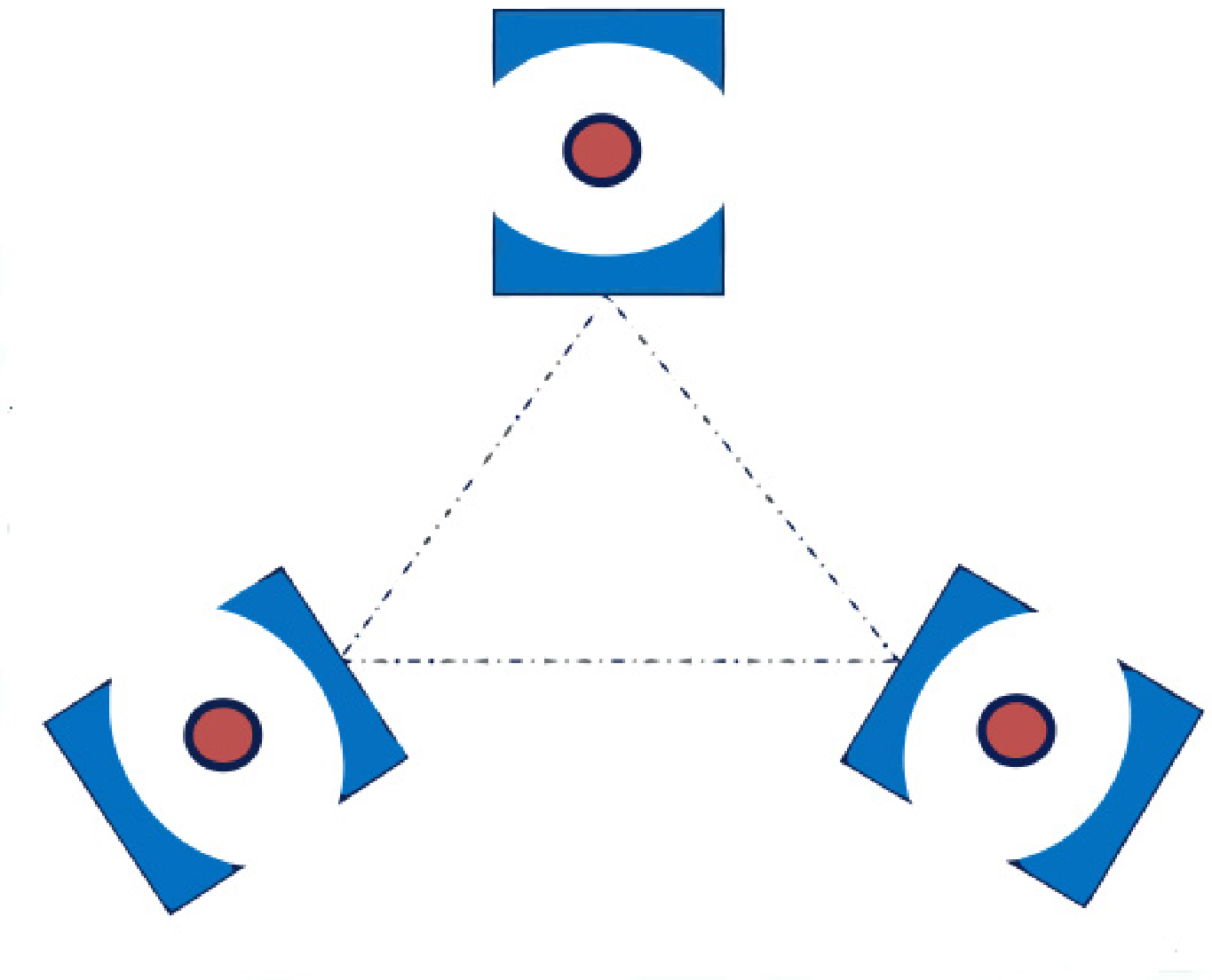}
\caption{} \label{Fig1}
\end{figure}
\newpage
\textbf{Figure Captions}
\itemize{}
\item Fig. 2. $E_{_{GHZ}}\left(\hat{\rho}(0, 2;
t)\right)$ and the related average number of photons ($\mathrm{<n_{c}>}$) for $c=1$, $g=30$ and $\Delta=0$.
\begin{figure}
\centering
\includegraphics[width=445 pt]{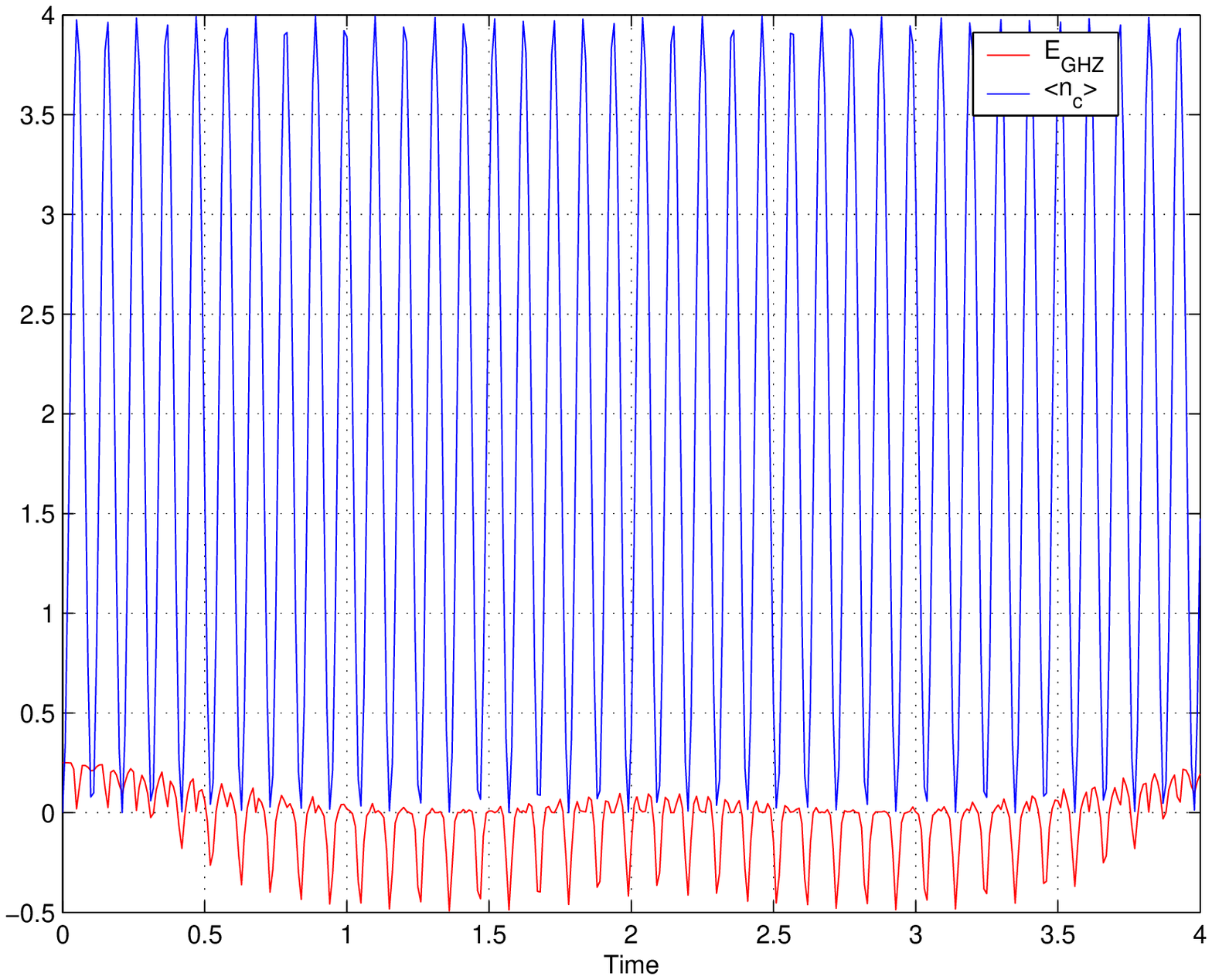}
\caption{} \label{Fig1}
\end{figure}
\newpage
\textbf{Figure Captions}
\itemize{}
\item Fig. 3. $E_{_{GHZ}}\left(\hat{\rho}(0, 2;
t)\right)$ and the related average number of photons ($\mathrm{<n_{c}>}$) for $c=1$, $g=30$ and $\Delta=-500$.
\begin{figure}
\centering
\includegraphics[width=445 pt]{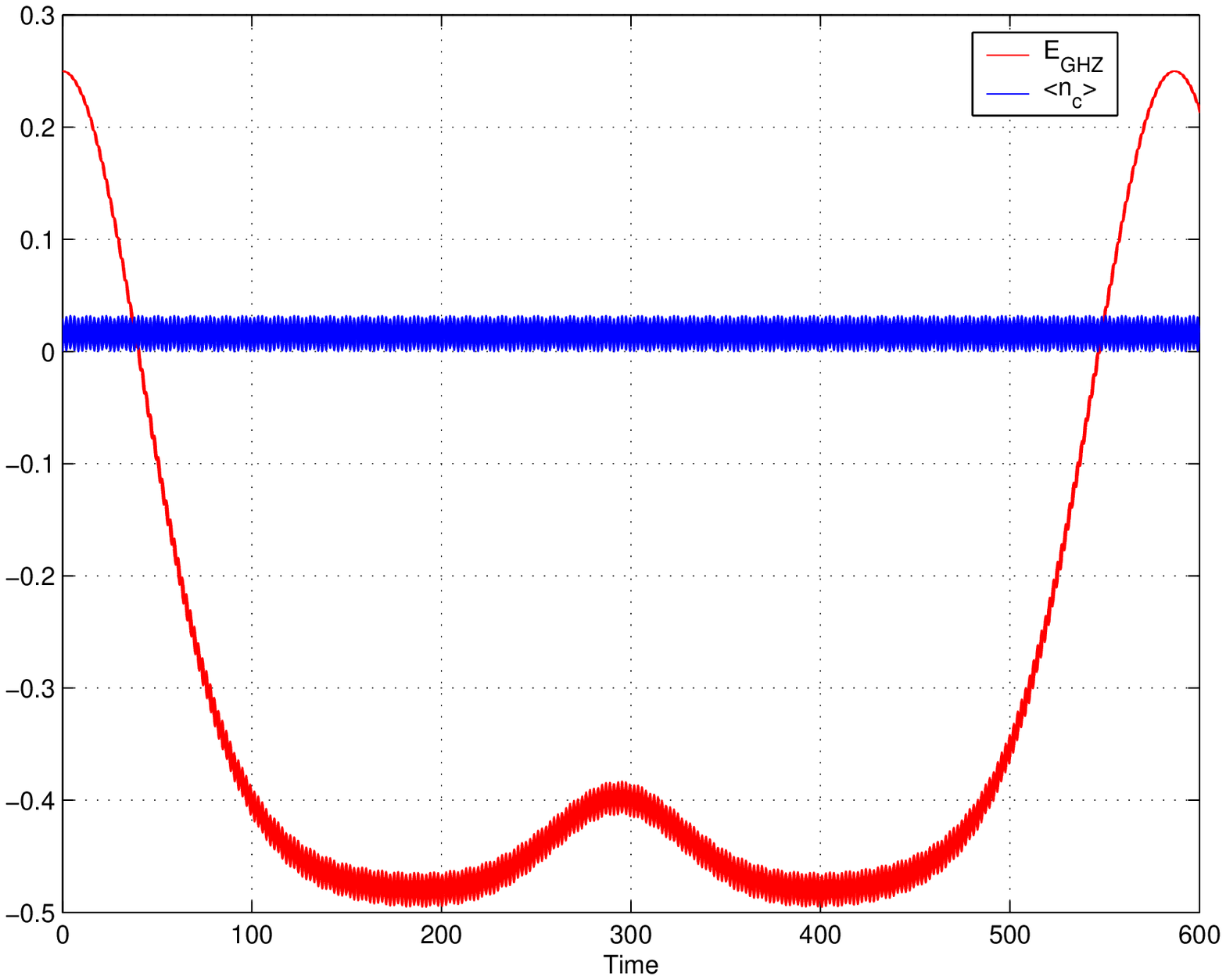}
\caption{} \label{Fig1}
\end{figure}
\newpage
\textbf{Figure Captions}
\itemize{}
\item Fig. 4. $E_{_{W}}\left(\hat{\rho}(\pi, 0.01; t)\right)$ and the related average number of photons ($\mathrm{<n_{c}>}$) for $c=1$, $g=30$ and $\Delta=0$.
\begin{figure}
\centering
\includegraphics[width=445 pt]{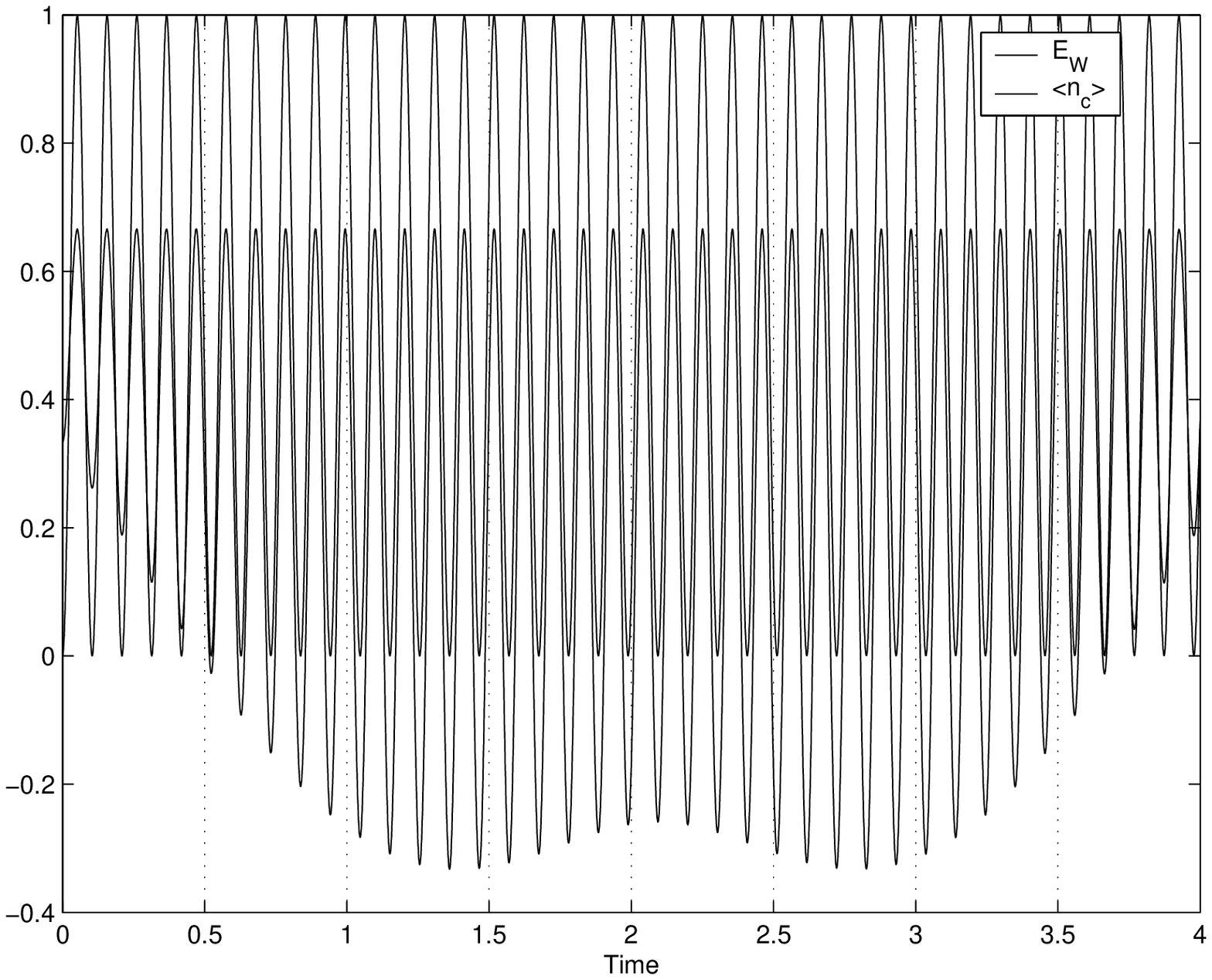}
\caption{} \label{Fig1}
\end{figure}
\newpage
\textbf{Figure Captions}

\item Fig. 5. $E_{_{W}}\left(\hat{\rho}(\pi, 0.01; t)\right)$ and the related average number of photons ($\mathrm{<n_{c}>}$) for $c=1$, $g=30$ and $\Delta=-500$.
\begin{figure}
\centering
\includegraphics[width=445 pt]{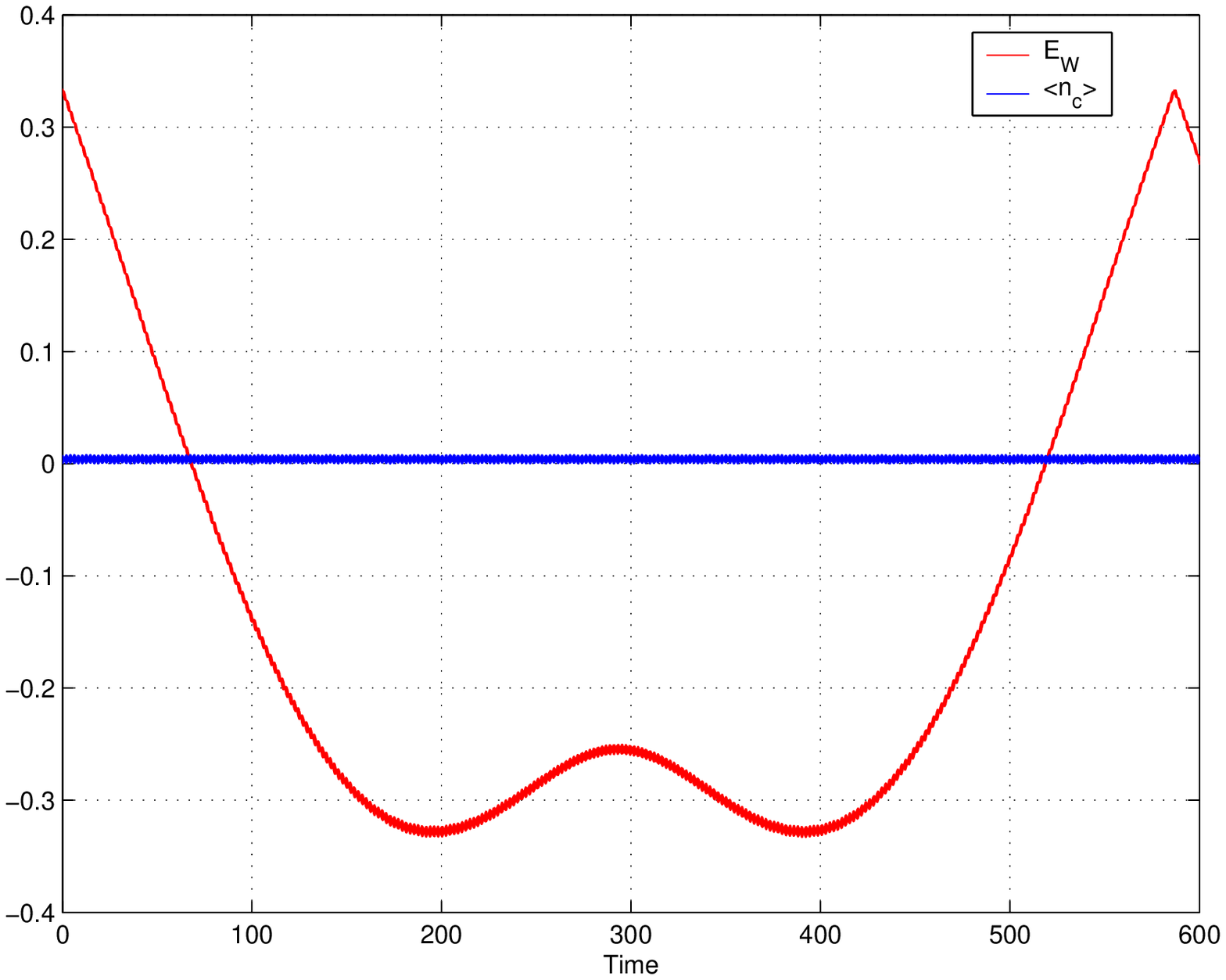}
\caption{} \label{Fig1}
\end{figure}
\newpage
\textbf{Figure Captions}

\item Fig. 6. $E_{_{GHZ}}\left(\hat{\rho}_{_{diss}}(0, 2; t)\right)$ for $c=1$, $g=30$, $\Delta=0$, $\gamma_{e}=0.001$ and $\gamma_{c}=0$.
\begin{figure}
\centering
\includegraphics[width=445 pt]{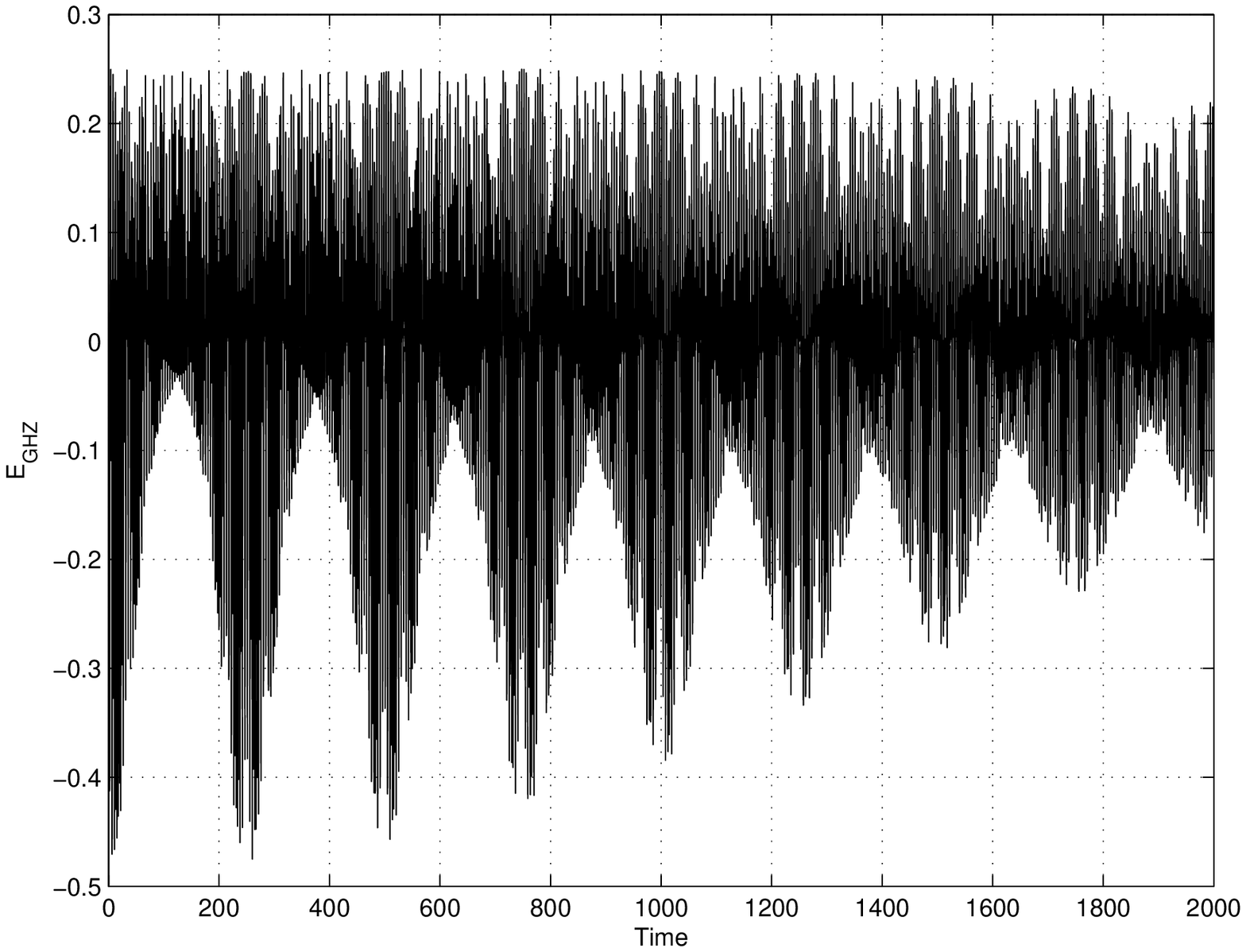}
\caption{} \label{Fig1}
\end{figure}
\newpage
\textbf{Figure Captions}

\item Fig. 7. $E_{_{GHZ}}\left(\hat{\rho}_{_{diss}}(0, 2; t)\right)$ for $c=1$, $g=30$, $\Delta=-500$, $\gamma_{e}=0.001$ and $\gamma_{c}=0$.
\begin{figure}
\centering
\includegraphics[width=445 pt]{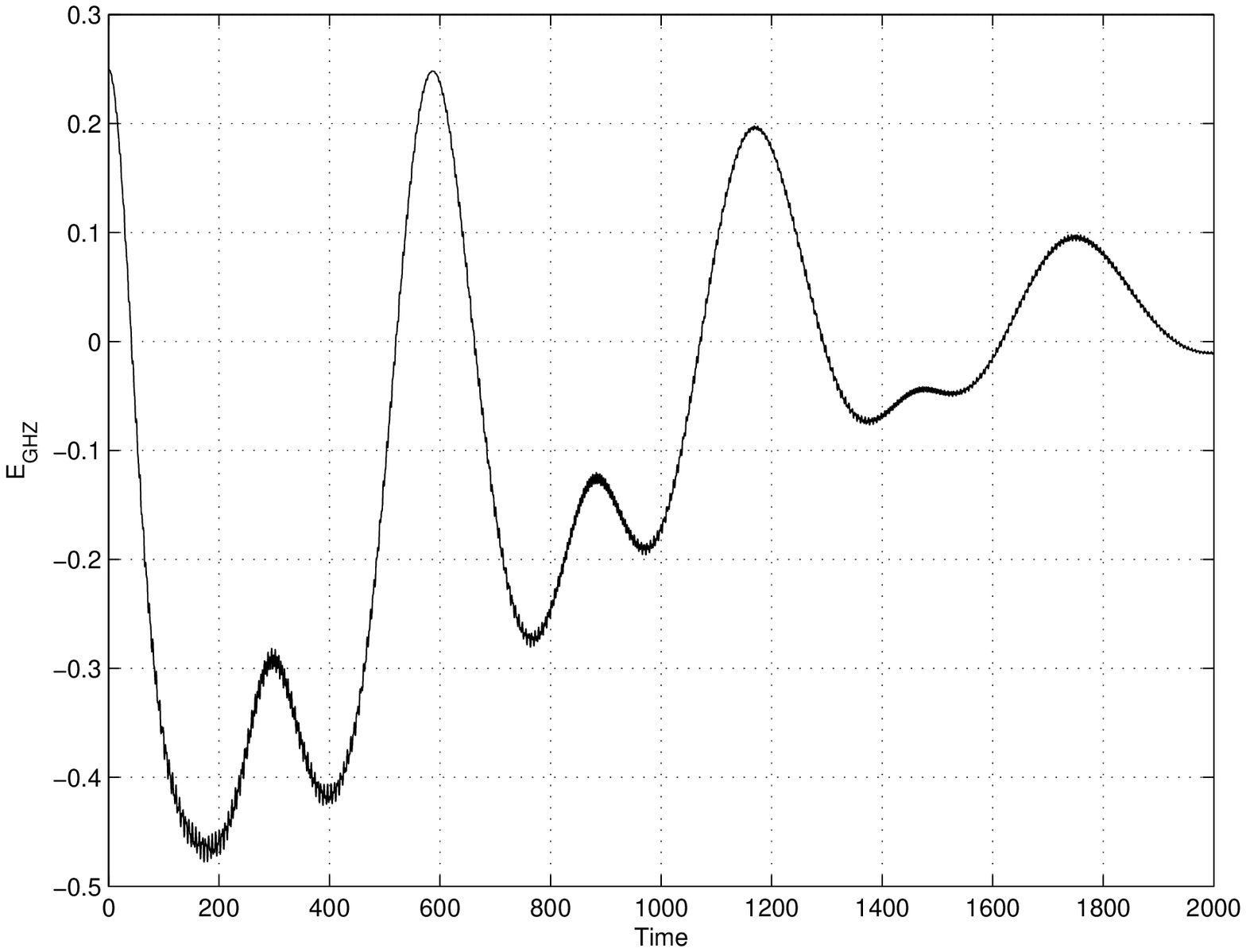}
\caption{} \label{Fig1}
\end{figure}
\newpage
\textbf{Figure Captions}

\item Fig. 8. $E_{_{GHZ}}\left(\hat{\rho}_{_{diss}}(0, 2; t)\right)$ for $c=1$, $g=30$, $\Delta=0$, $\gamma_{e}=0.001$ and $\gamma_{c}=0.05$. The destruction of the generated entangled coherent GHZ-state is considerable in comparison to the case that the photon losses in the cavities are not included (see Fig. 6).
\begin{figure}
\centering
\includegraphics[width=445 pt]{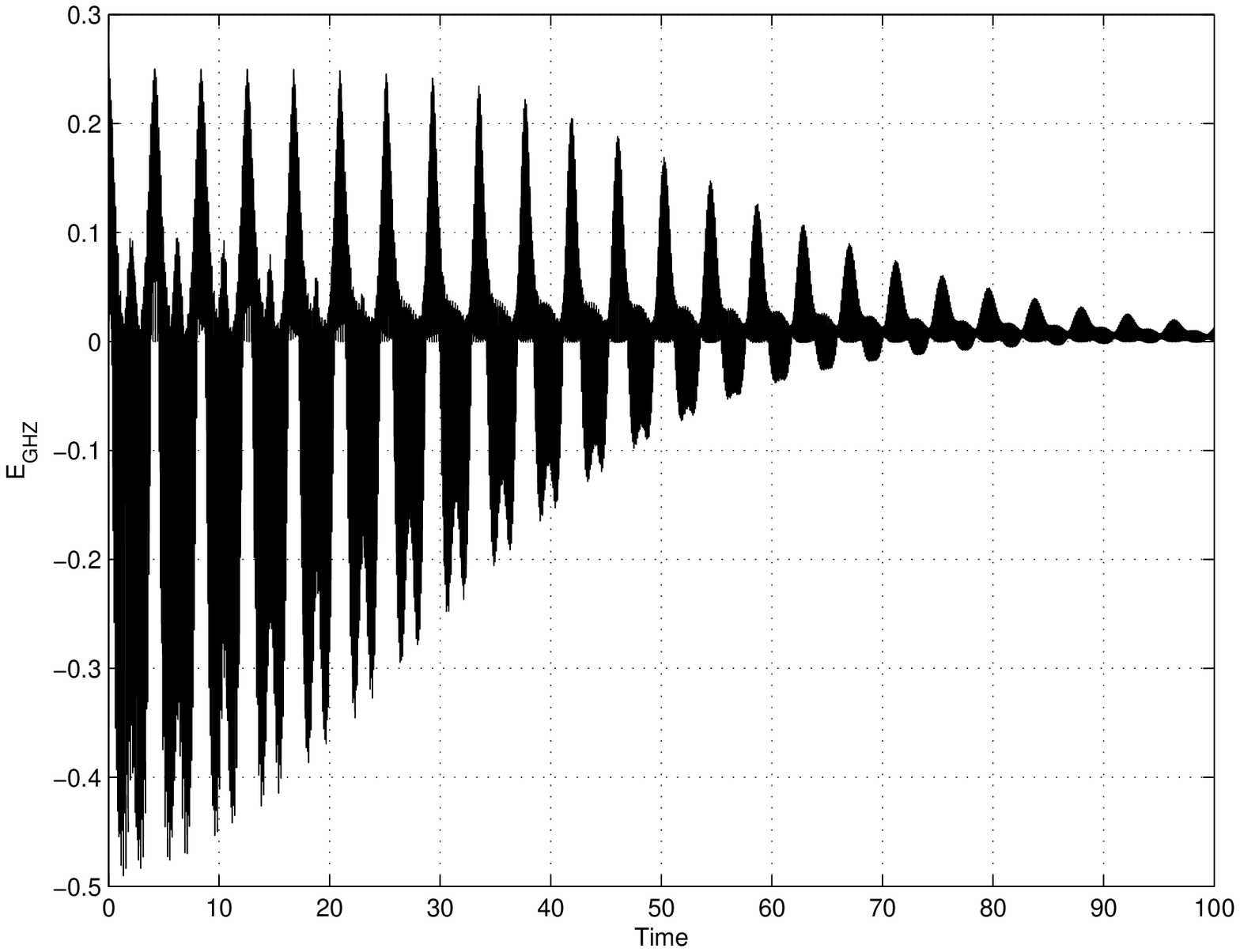}
\caption{} \label{Fig1}
\end{figure}
\newpage
\textbf{Figure Captions}

\item Fig. 9. $E_{_{GHZ}}\left(\hat{\rho}_{_{diss}}(0, 2; t)\right)$ for $c=1$, $g=30$, $\Delta=-500$, $\gamma_{e}=0.001$ and $\gamma_{c}=0.05$. For this case, obviously the robustness of entanglement due to the photon losses in the cavities is considerably more than one in the resonance case.
\begin{figure}
\centering
\includegraphics[width=445 pt]{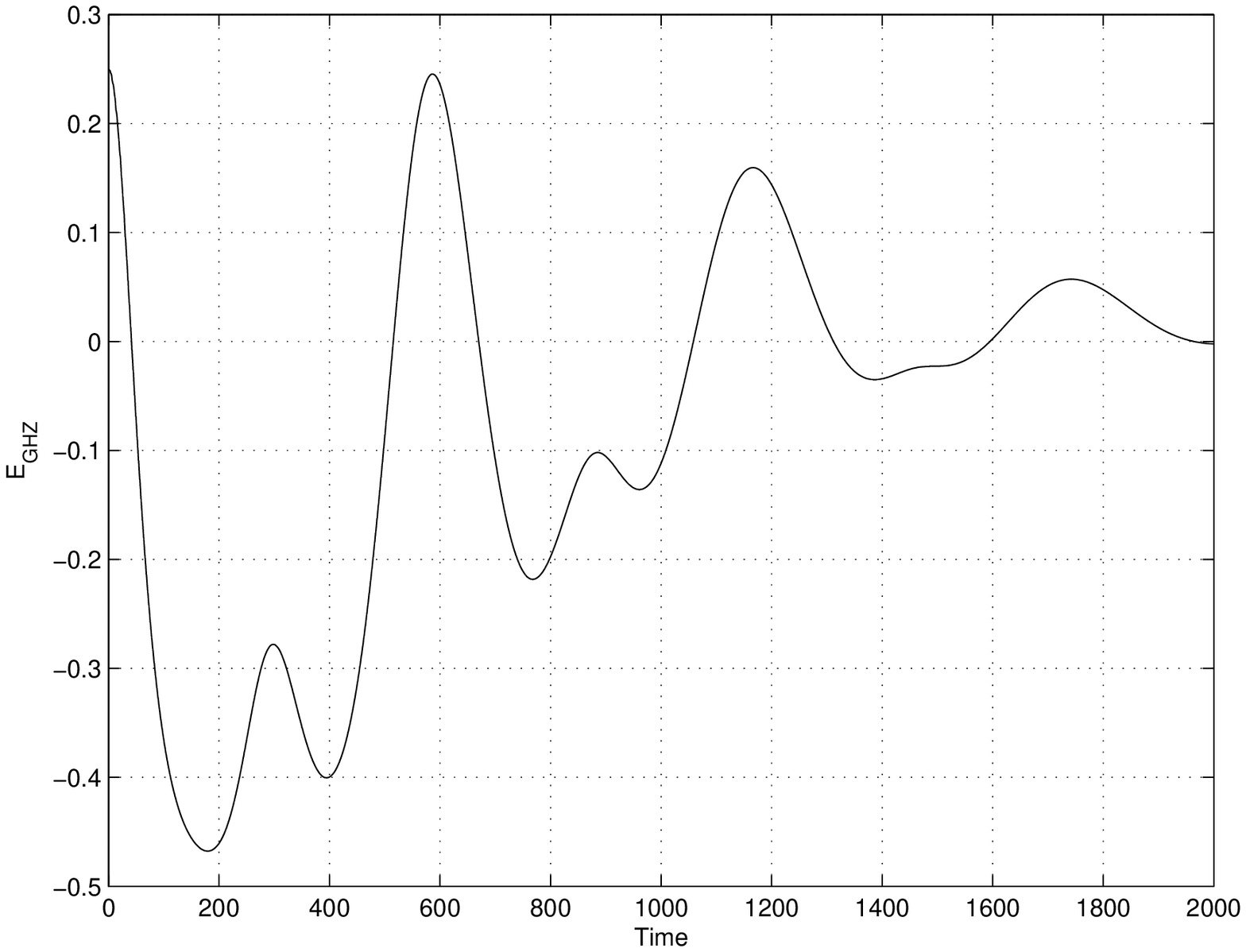}
\caption{} \label{Fig1}
\end{figure}
\newpage
\textbf{Figure Captions}

\item Fig. 10. $E_{_{W}}\left(\hat{\rho}_{_{diss}}(\pi, 0.01; t)\right)$ for $c=1$, $g=30$, $\Delta=0$, $\gamma_{e}=0.001$ and $\gamma_{c}=0$.
\begin{figure}
\centering
\includegraphics[width=445 pt]{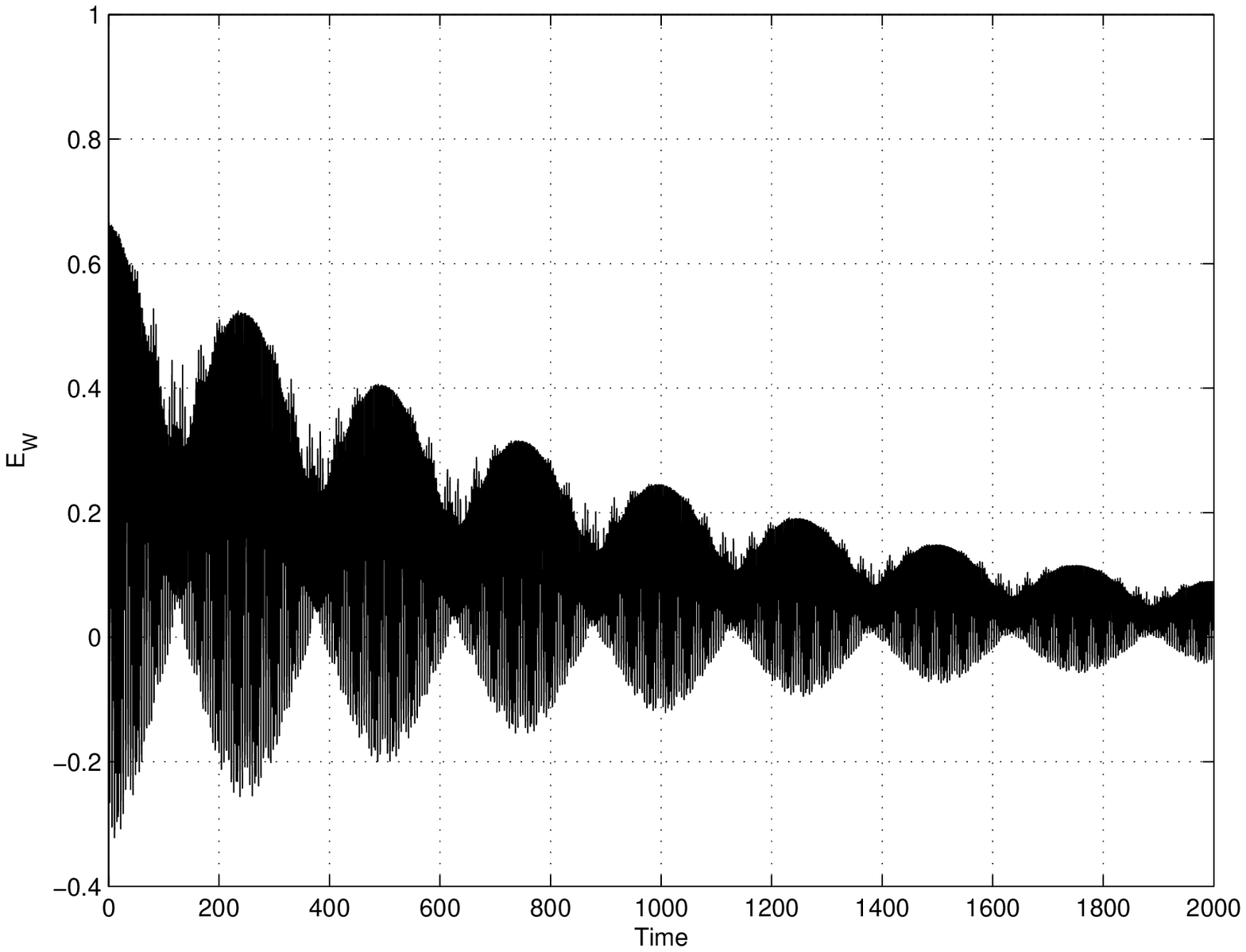}
\caption{} \label{Fig1}
\end{figure}
\newpage
\textbf{Figure Captions}

\item Fig. 11. $E_{_{W}}\left(\hat{\rho}_{_{diss}}(\pi, 0.01; t)\right)$ for $c=1$, $g=30$, $\Delta=-500$, $\gamma_{e}=0.001$ and $\gamma_{c}=0$.
\begin{figure}
\centering
\includegraphics[width=445 pt]{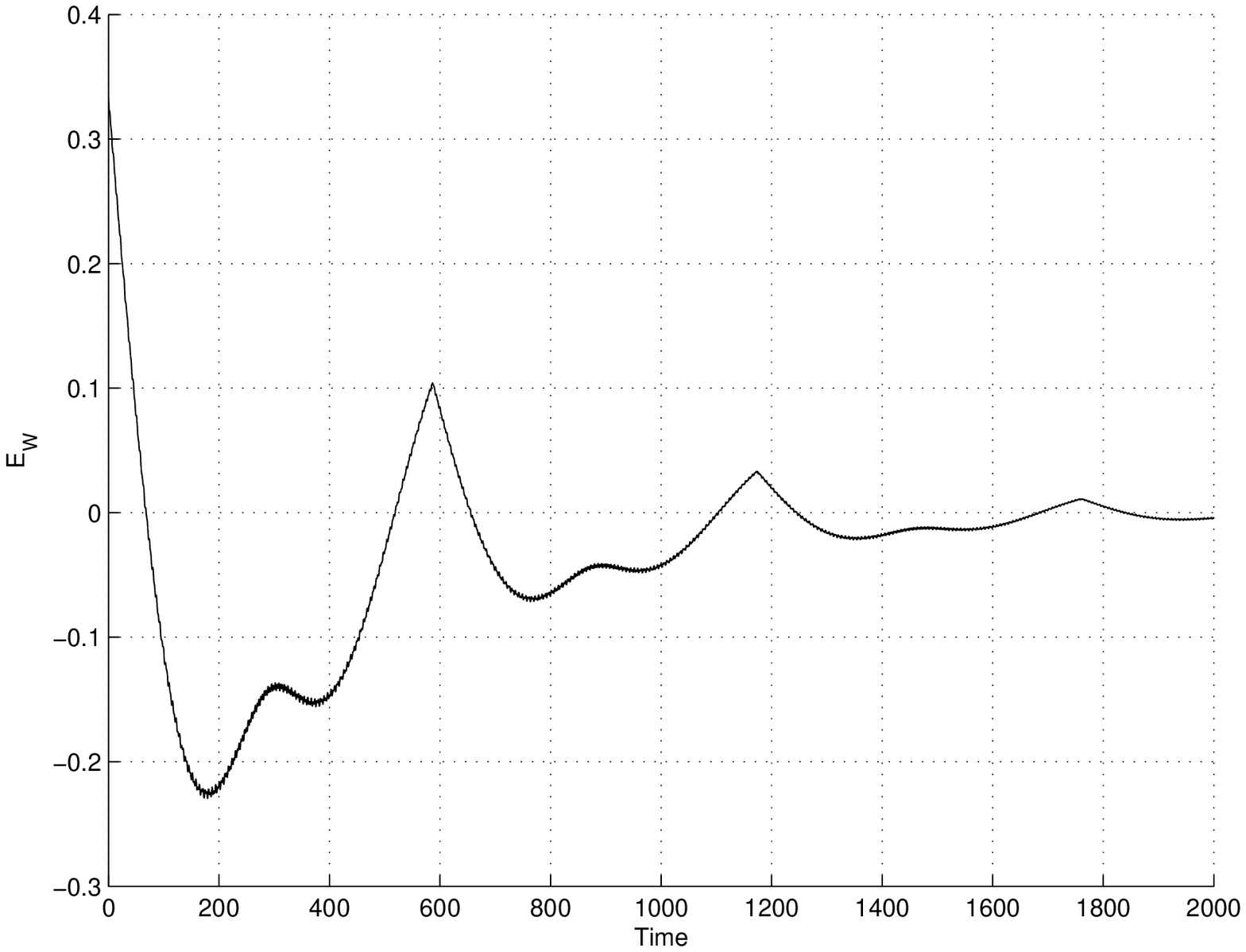}
\caption{} \label{Fig1}
\end{figure}
\newpage
\textbf{Figure Captions}

\item Fig. 12. $E_{_{W}}\left(\hat{\rho}_{_{diss}}(\pi, 0.01; t)\right)$ for $c=1$, $g=30$, $\Delta=0$, $\gamma_{e}=0.001$ and $\gamma_{c}=0.05$. Obviously, with these parameters, the destruction rate of the entanglement of coherent W-state due to the photon losses in the cavities is even more than one of the coherent GHZ-state as shown in Fig. 8.
\begin{figure}
\centering
\includegraphics[width=445 pt]{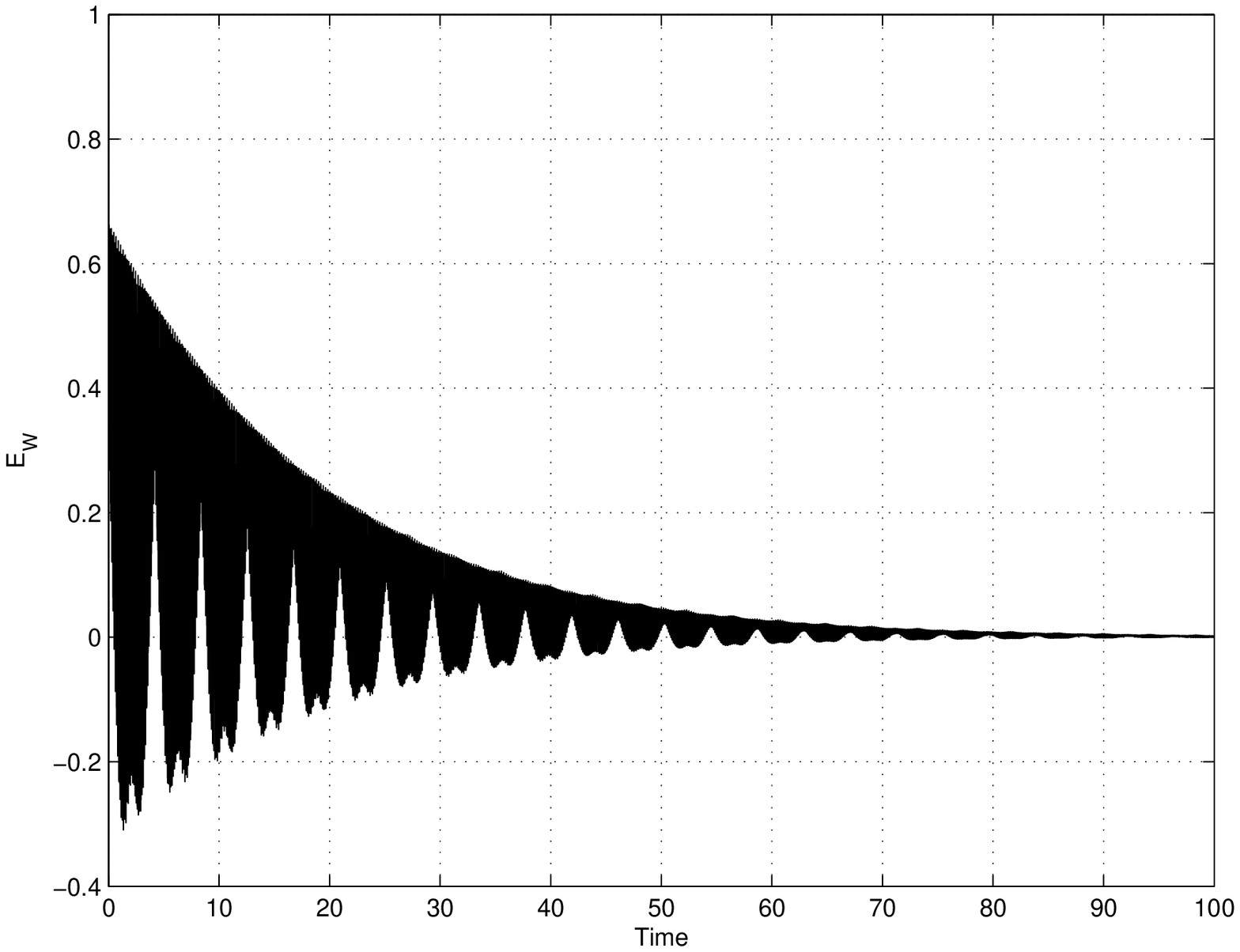}
\caption{} \label{Fig1}
\end{figure}
\newpage
\textbf{Figure Captions}

\item Fig. 13. $E_{_{W}}\left(\hat{\rho}_{_{diss}}(\pi, 0.01; t)\right)$ for $c=1$, $g=30$, $\Delta=-500$, $\gamma_{e}=0.001$ and $\gamma_{c}=0.05$. It is clearly observed that the robustness of entanglement of coherent W-state due to the photon losses in the cavities for nonresonance regime is considerably more than one for the resonance regime as depicted in Fig. 12.
\begin{figure}
\centering
\includegraphics[width=445 pt]{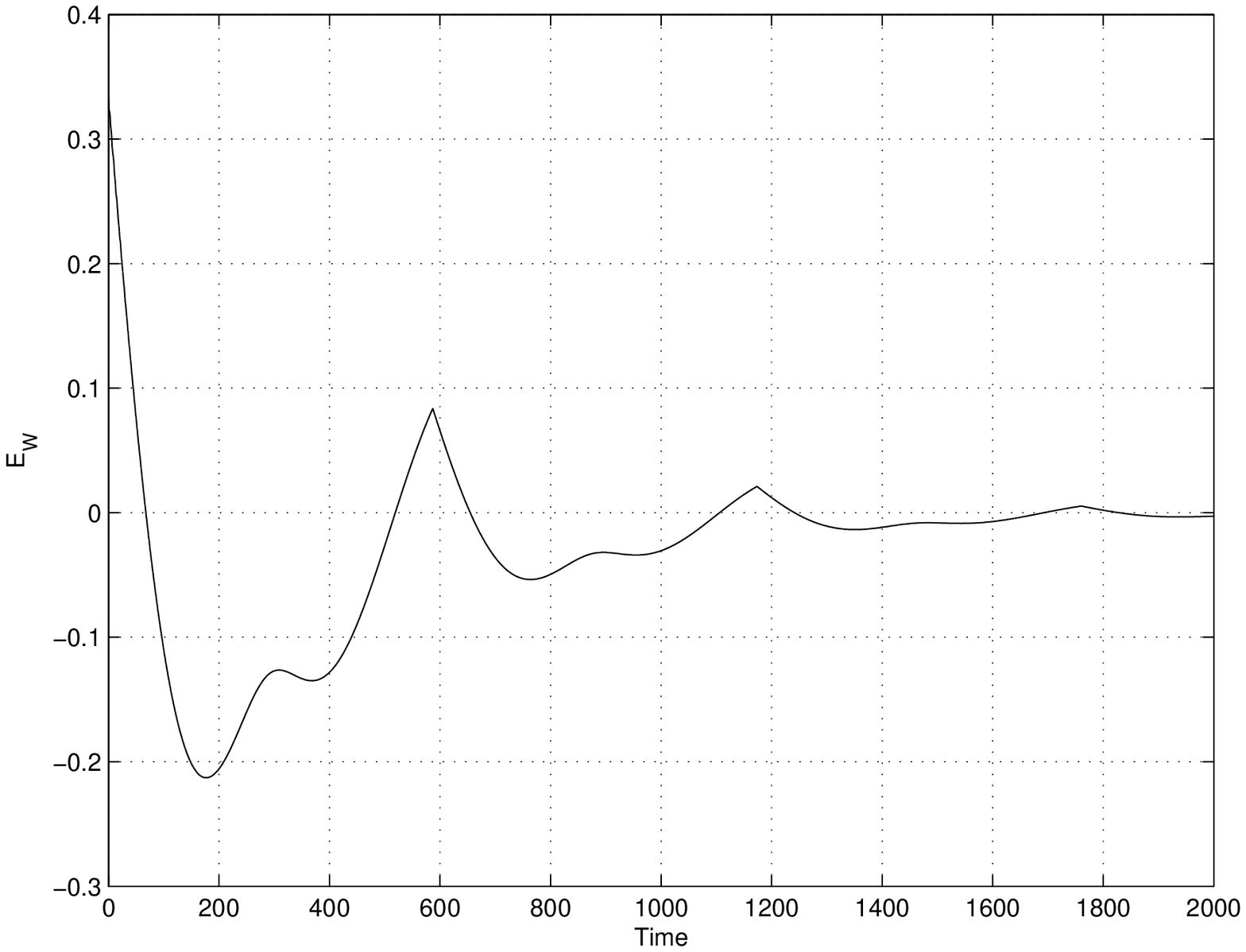}
\caption{} \label{Fig1}
\end{figure}

\newpage
\begin{thebibliography}{99}

\bibitem{Cirac1}
J. I. Cirac, A. K. Ekert, S. F. Huelga, and C. Macchiavello, Phys. Rev. A 59, 4249 (1999).

\bibitem{har}
M. Hartmann, F.G.S.L. Brandao, M.B. Plenio, Nat. Phys. 2, 849 (2006).

\bibitem{dim}
D. G. Angelakis, M. F. Santos and S. Bose, Phys. Rev. A 76, 031805(R) (2007).

\bibitem{lam}
L. Lamata, D. R. Leibrandt, I. L. Chuang, J. I. Cirac, M. D. Lukin, V. Vuletic and S. F. Yelin, Phys. Rev. Lett. 107, 030501 (2011).

\bibitem{ogd}
C.D. Ogden, E.K. Irish, M.S. Kim, Phys. Rev. A 78, 063805 (2008).

\bibitem{li}
P. B. Li, Y. Gu, Q. H. Gong, and G. C. Guo, Phys. Rev. A 79, 042339 (2009).

\bibitem{uti}
B. F. C. Yabu-uti, J. A. Roversi, Quantum Inf Process 12, 189 (2013).

\bibitem{Cirac2}
J. I. Cirac, P. Zoller, H. J. Kimble, and H. Mabuchi, Phys. Rev. Lett. 78, 3221 (1997).

\bibitem{henn}
K. Hennessy, A. Badolato, M. Winger, D. Gerace, M. Atature, S. Gulde, S. Falt, E.L. Hu, A. Imamoglu, Nature 445, 896 (2007).

\bibitem{khit}
G. Khitrova, H.M. Gibbs, M. Kira, W. Kochs, A. Scherer, Nat. Phys. 2, 81 (2006).

\bibitem{zho}
Z.R. Zhong, Opt. Commun. 283, 1972 (2010).

\bibitem{feng}
X.L. Feng, Z.M. Zhang, X.D. Li, Phys. Rev. Lett. 90, 217902 (2003).

\bibitem{pell}
T. Pellizzari, Phys. Rev. Lett. 79, 5242 (1997).

\bibitem{sera}
A. Serafini, S. Mancini, S. Bose, Phys. Rev. Lett. 96, 010503 (2006).

\bibitem{bell}
B. Bellomo, G. Compagno, R. Lo Franco, A. Ridolfo and S. Savasta, Phys. Scr. 2011, 014004 (2011).

\bibitem{gre}
D. M. Greenberger, M. Horne, and A. Zeilinger, in Bell's
Theorem, Quantum Theory, and Conceptions of the Universe,
edited by M. Kafatos (Kluwer, Dordrecht, 1989).

\bibitem{pan}
[2] J. W. Pan, D. Bouwmeester, M. Daniell, H. Weinfurter, and
A. Zeilinger, Nature (London) 403, 515 (2000).

\bibitem{che}
R. Chen, L. Shen, Phys. Lett. A 375, 3840 (2011).

\bibitem{liw}
W.-A. Li and L.-F. Wei, OPTICS EXPRESS 20, 13440 (2012).

\bibitem{lu}
X.-Y. Lu, L.-G. Si, X.-Y. Hao, X. Yang, Phys. Rev. A 79, 052330 (2009)

\bibitem{vid}
W. Duer, G. Vidal, and J. I. Cirac, Phys. Rev. A 62, 062314 (2000).

\bibitem{kem}
J. Kempe, Phys. Rev. A 60, 910 (1999); A. V. Thapliyal, ibid.
59, 3336 (1999); D. Gottesman and I. L. Chuang, Nature
(London) 402, 390 (1999); M. A. Nielsen and I. L. Chuang,
Quantum Computation and Quantum Information (Cambridge University Press, Cambridge, 2000).

\bibitem{san}
B. C. Sanders, Phys. Rev. A 45, 6811 (1992).

\bibitem{lun}
A. P. Lund, T. C. Ralph, and H. L. Haselgrove, Phys. Rev. Lett. 100, 030503 (2008).

\bibitem{an1}
N. B. An, Phys. Lett. A 373, 1701 (2009).

\bibitem{mar}
P. Marek and J. Fiurasek, Phys. Rev. A 82, 014304 (2010).

\bibitem{an2}
H. Jeong, N. B. An, Phys. Rev. A 74, 022104 (2006).

\bibitem{an3}
N. B. An, Phys. Rev. A 69, 022315 (2004).

\bibitem{acin}
A. Acin, D. Brus, M. Lewenstein, A. Sanpera, Phys. Rev. Lett. 87, 040401 (2001).

\bibitem{bou}
M. Bourennance, M. Eibl, C. Kurtsiefer, S. Gaertner, H. Weinfurter, O. Guhne, P. Hyllus, D. Brus, M. Lewenstein,
A. Sanpera, Phys. Rev. Lett. 92, 087902 (2004).

\bibitem{dav}
L. Davidovich, M. Brune, J. M. Raimond, and S. Haroche, Phys. Rev. A 53, 1295 (1996).

\bibitem{fon}
K. M. Fonseca Romero, M. C. Nemes, J. G. Peixoto de Faria, A. N. Salgueiro, and A.
F. R. de Toledo Piza, Phys. Rev. A 58, 3205 (1998).

\bibitem{bar}
X. Wang and B. C. Sanders, Phy. Rev. A 65, 012303 (2001).

\bibitem{zhe}
S.B. Zheng, G.C. Guo, Phys. Rev. Lett. 85, 2392 (2000).

\bibitem{maj}
J. Majer et al., Nature 449, 443 (2007).

\bibitem{gar}
C. Gardiner and P. Zoller, Quantum Noise, Springer (2004).

\bibitem{lou}
W. H. Louisell and W. H. Marburger, IEEE J. Quantum Electron. 3, 348 (1967).

\bibitem{ple}
M. J. Hartmann, J. Prior, S. R. Clark, and M. B. Plenio, Phys. Rev. Lett. 102, 057202 (2009).

\end{thebibliography}
\end{document}